\documentclass[a4paper]{jpconf}
\usepackage{graphicx}
\usepackage{amsmath}
\usepackage{cite}
\usepackage{comment}

\newcommand{\slide}[1]{\begin{frame} \frametitle{\small #1}}
\newcommand{\plainslide}[1]{\begin{frame}[plain] \frametitle{ #1}}

\newcommand{\bc}{\begin{columns}}
\newcommand{\ec}{\end{columns}}

\newcommand{\vecr}{{\vec r}}
\newcommand{\vecrp}{\vec{r}\mkern2mu\vphantom{r}'}
\newcommand{\vecR}{{\vec R}}
\newcommand{\vecK}{{\vec K}}
\newcommand{\veck}{{\vec k}}

\newcommand{\vecRp}{{\vec R'}}

\begin{document}
\title{Reaction Theory and Advanced CDCC}

\author{A.M. Moro$^1$, J.~Casal$^2$, Jin Lei$^3$, M.~G\'omez-Ramos$^{1,}$\footnote[4]{Present address: Institut f\"ur Kernphysik, Technische Universit\"at Darmstadt, D-64289 Darmstadt, Germany.}}
\address{$^1$ Departamento de FAMN, Universidad de Sevilla, Apartado 1065, 41080 Sevilla, Spain.}
\address{$^2$ Dipartimento di Fisica e Astronomia ``G. Galilei'' and INFN Sezione di Padova, I-35131 Padova, Italy.}
\address{$^3$ Institute of Nuclear and Particle Physics, and Department of Physics and Astronomy, Ohio University, Athens, Ohio 45701, USA.}

\ead{moro@us.es}

\begin{abstract}
The Continuum-Discretized Coupled-Channels (CDCC) has been successfully employed to describe elastic and breakup of nuclear reactions induced by weakly bound projectiles. In this contribution, we review some other, less widespread applications of the CDCC wavefunction, some of them in combination with other reaction formalisms, which are being currently employed in the analysis of reactions involving three or more fragments in the initial or final state. 
\end{abstract}

\section{Introduction: reminder of the CDCC method}
The CDCC method was first introduced by G.\ Rawitscher \cite{Raw74} and later refined and fully implemented by the Pittsburgh-Kyushu collaboration \cite{Yah86,Aus87} to describe the effect of the breakup channels on the elastic scattering of deuterons. Denoting the reaction by $a+A$, with $a=b+x$ (referred hereafter as the {\it core} and {\it valence} particles, respectively), the method assumes that the many-body reaction can be reduced to an effective three-body problem described by the effective Hamiltonian
\begin{equation}
\label{eq:Heff}
H= H_\text{proj} + \hat{T}_{\vecR}   + U_{bA}(\vecr_{bA}) + U_{xA}(\vecr_{xA}) ,
\end{equation}
with $H_\mathrm{proj}=\hat{T}_{\vecr}+V_{bx}$ the projectile internal Hamiltonian,   $\hat{T}_{\vecr}$  and  $\hat{T}_{\vecR}$ are kinetic energy operators, $V_{bx}$ the inter-cluster interaction and $U_{bA}$ and $U_{xA}$ are the core-target and valence-target optical potentials (complex in general) describing the elastic scattering of the corresponding $b+A$ and $x+A$ sub-systems, at the same energy per nucleon of  the  projectile. 
In the  CDCC  method,  the  three-body wave function of the system is expanded in terms of the eigenstates of the Hamiltonian $H_\mathrm{proj}$ including both bound and unbound states. Since the latter form a continuum, a procedure of discretization is applied, consisting in approximating this continuum by a finite and discrete set of square-integrable functions. In actual calculations, this continuum must be truncated in excitation energy and limited to a 
finite number of partial waves $\ell$ associated to the relative co-ordinate $\vecr$. Normalizable states representing the continuum should be obtained for each $\ell,j$ values. Two main  types of discretization methods are commonly used. One is the {\it the pseudo-state method}, in which the $b+x$ Hamiltonian is diagonalized in a basis of square-integrable functions, such as Gaussians \cite{Kaw86a} or transformed harmonic oscillator functions \cite{Mor09b}. Negative eigenvalues correspond to the bound states of the systems, whereas positive eigenvalues are regarded as a finite representation of the continuum. The other is the {\it binning method}, in which  normalizable states are obtained by constructing  wave packets ({\it bins}) by linear superposition of the actual continuum states over a certain energy interval \cite{Aus87}.

Assuming a single-bound state for simplicity of the notation, the three-body CDCC wavefunction is then written as: 
\begin{align}
\label{eq:cdccwf}
\Psi^\mathrm{CDCC} (\vecr,\vecR) & = 
\phi_0 (\vecr)  \chi_0(\vecK_0,\vecR) + 
  \sum_{n}^{N} \phi_{n}(k_{n},\vecr) \chi_{n}(\vecK_{n},\vecR) 
\end{align}
where $n$ is a discrete index for the discretized continuum states,
$K$ the wavenumber associated  to the projectile-target relative motion 
and $k$ the nominal wavenumber of the bound continuum bins. 
The unknown functions $\{\chi_n(K_0,\vecR)\}$  ($n=0,\ldots,N$) are obtained by inserting $\Psi^\mathrm{CDCC}$ into the Schr\"odinger equation, giving rise to a system of coupled equations which must be solved subject to the boundary asymptotic condition
\begin{equation}
\label{eq:cdcc_asym}
\chi^{(+)}_{n}(\vecK, \vecR)  \xrightarrow{R\gg} e^{i \vecK \cdot \vecR} \delta_{n,0} +  f_{n,0}(\theta) \frac{e^{i KR }}{R} 
\end{equation}
where $f_{n,0}(\theta)$ is the scattering amplitude corresponding to the transition to the state $n$. The differential cross section for the population of this state is 
\begin{equation}
 \left (  \frac{d\sigma(\theta)}{d\Omega} \right )_{0\rightarrow n} = \frac{K_n}{K_0}  | f_{n,0}(\theta) |^2  \, .
\end{equation}

\section{Recent extensions and applications of the CDCC method \label{sec:cdcc_extend}}
In its original formulation, the CDCC method was restricted to two-body projectiles and ignored any possible excitations of the target and of the constituent fragments of the projectile.  Although these excitations are in principle taken into account, in an effective way, by the imaginary part of the fragment-target optical potentials, there are situations in which they may need to be explicitly taken into account.  Furthermore, the two-body picture may be inadequate for some nuclei such as, for example, for  Borromean systems (e.g.~$^{6}$He, $^{9}$Be, $^{11}$Li).  Suitable extensions of the CDCC formalism aimed at overcoming these limitations  have been developed since the early days of the method. For example, collective excitations of the target nucleus can be incorporated using an augmented modelspace which includes some excited states of this nucleus and  replacing the fragment-target interactions $U_{bA}(\vecr_{bA})$ and  $U_{xA}(\vecr_{xA})$ [c.f.\ Eq.~(\ref{eq:Heff})] by deformed potentials. This was first done by Yahiro {\it et al.} \cite{Yah86} for the case of deuteron breakup, although the calculations were restricted to $p$-$n$ $s$-waves.  The problem has been also revisited recently  \cite{Gom17a} and extended to other weakly bound projectiles and higher partial waves. In the case of deuteron reactions, the computed elastic scattering and inelastic scattering (i.e.~target excitation) cross sections were found to reproduce very well those found with Faddeev/AGS calculations \cite{Gom17a}. 

To account for possible excitations of the projectile fragments, an extended version of the CDCC method (coined XCDCC) has been also formulated in recent years \cite{Sum06,Die14}. In the XCDCC method, the effective Hamiltonian (\ref{eq:Heff}) is modified by using deformed fragment-target potentials and by including the effect of the deformation of one of the fragments ($b$ or $x$) in the projectile Hamiltonian  $H_\text{proj}$ using, for instance, a particle-rotor model.  The method has been applied to reactions induced by $^{11}$Be \cite{Sum07,Mor12b,Die14,Pes17,Chen16a,Chen16b,Dip19} and $^{19}$C  \cite{Lay16}. The main conclusion of these calculations is that, in the case of light targets (such as protons or $^{12}$C), the dynamical excitation mechanism of the core nucleus plays a very important in the reaction, producing a sizable increase of the breakup cross sections \cite{Mor12b,Die14,Lay16}. For heavier targets, such as $^{64}$Zn \cite{Dip19} or  $^{197}$Au \cite{Pes17}, this dynamical excitation mechanism is very small, which is attributed to the fact that the breakup is dominated by the long-range  dipole Coulomb couplings tending to dissociate $x$ from $b$, which overwhelms the effect of the quadrupole and octupole collective couplings exciting collective modes of the core ($b$). Yet, the inclusion of the core deformation  may still affect significantly the $B(E\lambda$) strength between bound states of the projectile and this translates into important differences in the inelastic scattering cross sections for the population of these states  \cite{Pes17}. 


\section{Applications of the CDCC wavefunction to other formalisms}
In principle, the CDCC wavefunction provides only elastic and elastic breakup cross sections since these are the only components present in the asymptotic form (\ref{eq:cdcc_asym}), from which the scattering observables are computed. Yet, the full CDCC wavefunction (\ref{eq:cdccwf}) contains additional information on the reaction. It can be regarded as an accurate representation of the three-body  wavefunction of the system, at least for $b$-$x$ separations spanned by the discrete basis $\{ \phi_n (\vecr) \}$. As such, it can be combined with other reaction frameworks  which require some approximation of the three-body wavefunction within that region of the configuration space. Some recent examples are given in the following subsections. 

\subsection{Transfer reactions induced by weakly bound nuclei on deformed targets}
Theories of $(d,p)$  and $(p,d)$  reactions frequently rely on formalisms based on a transition amplitude that is dominated by the components of the total three-body scattering wave function where the spatial separation between the incoming neutron and proton is confined by the range of the $n$-$p$ interaction. This is where the CDCC wavefunction for the $d+A$ system in terms of $p$-$n$ states is at its best, so it be used for that purpose. The idea has been used by several authors (see e.g.~ \cite{Pan13} and \cite{Cha17}) to assess the validity of other, more approximate methods, such as the adiabatic approximation \cite{Ron70,JT74}.  


Additional dynamical effects can be studied with the extended versions of the CDCC wavefunction discussed in Sec.~\ref{sec:cdcc_extend}. For example, the extended CDCC wavefunction with target excitations can be used to investigate the effect of target excitation in transfer reactions. This idea has been recently applied to the  $^{10}\text{Be}(d,p)^{11}\text{Be}$ reaction at difference incident energies \cite{Gom17b}. For that, one may use the post-form  of the transition matrix for the process $A(d,p)B$, i.e.,
\begin{equation}
\label{eq:T}
{\cal T}_{dp}= \langle \chi^{(-)}_p \Phi_B | V_{pn} + U_{pA}-U_{pB} | \Psi^{(+)}_d \rangle  ,
\end{equation}
where $V_{pn}$, $U_{pA}$ are the proton-neutron and proton-target interactions, $U_{pB}$ is an auxiliary (and, in principle, arbitrary) potential for the $p$-$B$ system, $\Phi_B$  is the internal wave function of the residual nucleus $B$ and $\Psi^{(+)}_d$ is the exact total wave function corresponding  to an incident deuteron beam of kinetic energy $E_d$ and binding energy $\varepsilon_d$. This function can be suitably approximated by the extended CDCC wavefunction.
\begin{align}
\label{eq:Phi}
\Psi^{(+)}_d(\vecr,\vecR,\xi) & = \phi_d(\vecr) \Phi^{(0)}_{A} (\xi) \chi^{(+)}_{d,0}(\vecR)   
              + \phi_d(\vecr) \Phi^{(0)}_{A} (\xi) \chi^{(+)}_{d,2}(\vecR) \nonumber \\
              & + \sum_{i=1}^{N} \phi^{i}_{pn}(\vecr) \Phi^{(2)}_{A} (\xi) \chi^{(+)}_{i,0}(\vecR) 
              + \sum_{i=1}^{N} \phi^{i}_{pn}(\vecr) \Phi^{(2)}_{A} (\xi) \chi^{(+)}_{i,2}(\vecR),
\end{align}
where $ \{\phi_d, \phi^{i}_{pn} \}$ denote the deuteron ground state and discretized $p$-$n$ continuum states, $\Phi^{(I)}_{A} (\xi)$ the target wavefunction in the ground ($I=0$) or excited ($I=2$)  state and  $\{\chi^{(+)}_{i,I}(\vecR) \}$ the functions describing the projectile-target relative motion with the target in the state $I$ and the $p$-$n$ system in the state $i$. Thus,  the first two terms of Eq.~(\ref{eq:Phi}) describe, respectively, elastic and inelastic scattering with the deuteron remaining in its ground state. The third  and fourth terms describe deuteron breakup with respect to the target in its ground state or first excited state, respectively.  When inserted into Eq.~(\ref{eq:T}) this gives rise also to four terms, 
\begin{equation}
\label{eq:Tdecomp}
{\cal T}_{dp}= {\cal T}^\mathrm{el}_{dp} + {\cal T}^\mathrm{inel}_{dp} +{\cal T}^\mathrm{elbu}_{dp} + {\cal T}^\mathrm{inbu}_{dp} ,
\end{equation}
which may be interpreted as (I) {\it elastic transfer}, i.e.,  direct transfer from the deuteron ground state leaving the target in its ground state, (II) {\it inelastic transfer}, i.e., target excitation followed by one-neutron transfer, (III) {\it elastic breakup transfer}, i.e.,  deuteron breakup followed by transfer, leaving the target in the ground state and (IV)  {\it inelastic breakup transfer}, i.e., deuteron breakup, accompanied by target excitation, and followed by neutron transfer.

Likewise, the wave function $\Phi_B$ for a total angular momentum $J$ and projection $M$ will contain contributions from different $A$ states. Using the usual parentage decomposition, and ignoring antisymmetrization for simplicity, one may write
\begin{equation}
\label{eq:AB}
\Phi^{JM}_B(\vecr_{nA},\xi)= \sum_{I,l,j} [ \Phi^I_A(\xi) \otimes \phi_{l j }(\vecr_{nA}) ]_{JM} ,
\end{equation}
where $l$ and $j$ the orbital and total ($\vec{j}=\vec{l}+s$) angular momentum of the valence particle and  $\phi_{l j }(\vecr_{nA})$   is  a function describing the neutron-core relative motion. 

\begin{figure}[htbp]
\begin{minipage}{0.55\textwidth}
\includegraphics[width=16pc]{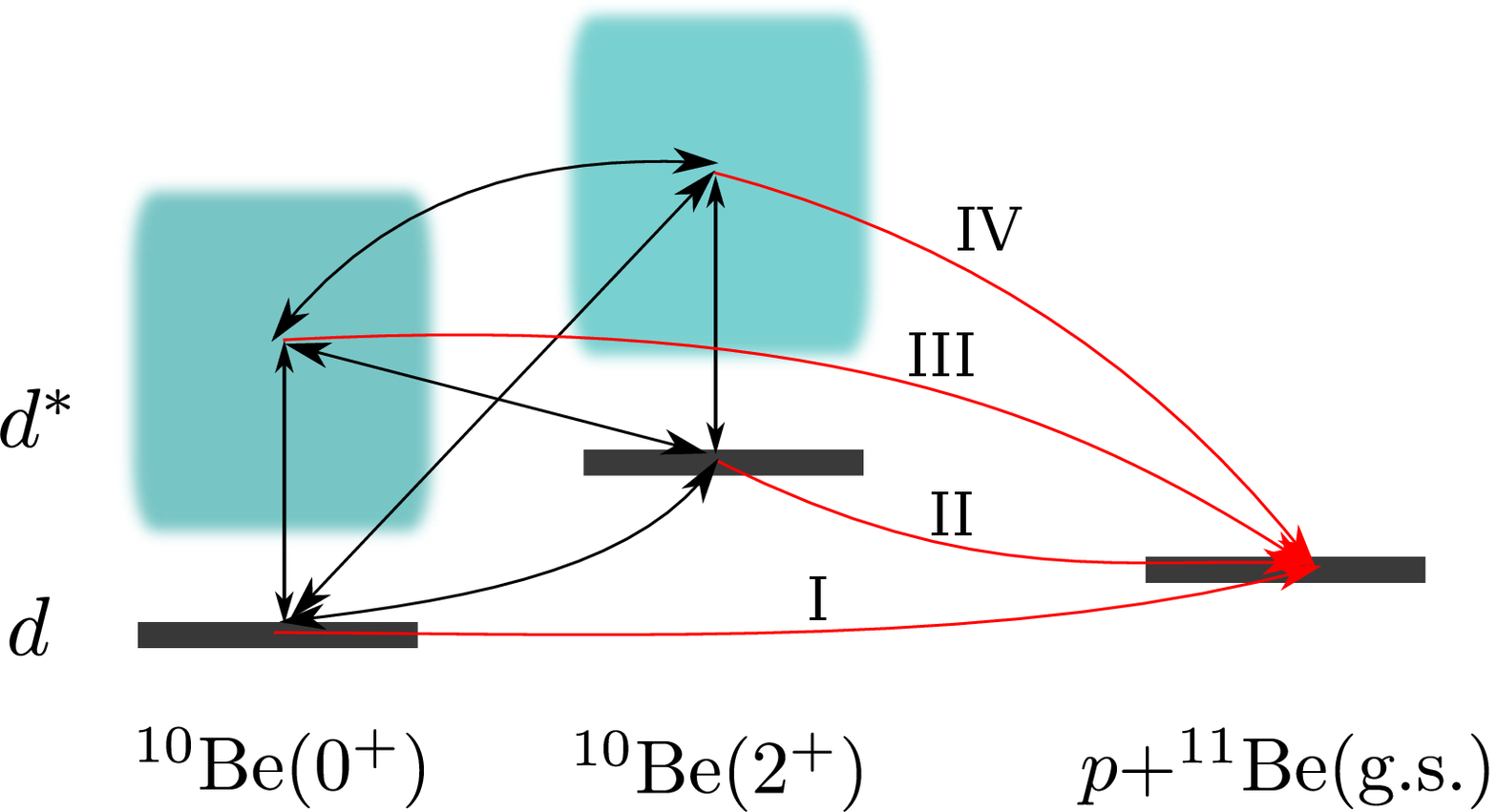}
\end{minipage}\hspace{0pc}%
\begin{minipage}{0.4\textwidth}
\includegraphics[width=12pc]{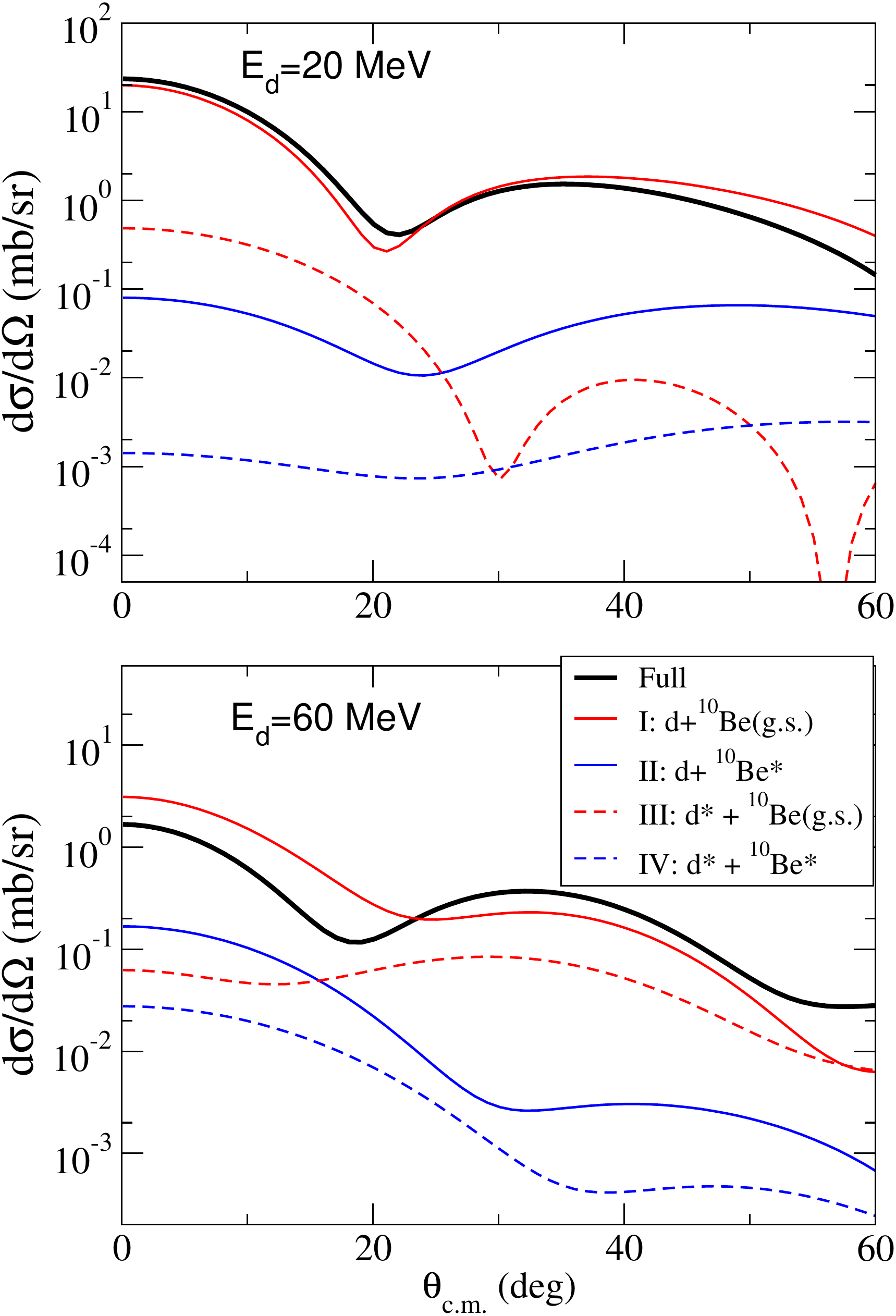}
\end{minipage} 
\caption{\label{fig:cdcc_tarx} Application of the CDCC method to the transfer reaction $^{10}$Be($d$,$p$)$^{11}$Be, including simultaneously the effects of deuteron breakup and target excitation. LHS:  Coupling scheme showing the different paths. RHS: Contributions of the different paths to the transfer angular distribution, for two incident energies of the deuteron. }
\end{figure}

A recent application of this formalism \cite{Gom17b} is shown  in Fig.~\ref{fig:cdcc_tarx}. On the left-hand-side we depict the coupling scheme, with the different paths contributing to the $^{10}\text{Be}(d,p)^{11}\text{Be}(\text{g.s.})$ reaction. Path I is the direct, one-step transfer in which the neutron is directly transferred from the deuteron g.s.\ to the $^{11}\text{Be}$ ground-state. This corresponds to the usual DWBA calculation. The other paths (II-IV) involve multistep processes proceeding via the $p$-$n$ and/or the target excited state.  On the right-hand side, we show the separate contribution of each of these paths to the transfer differential cross section, for two different deuteron incident energies. It is seen that both, deuteron breakup and target excitation, gain importance for the higher energy. Also, it is to be noted that the full calculation is the coherent sum of the different paths [c.f.\ Eq.~(\ref{eq:Tdecomp})] so interference effects appear. 

%

\subsection{Application to proton-induced knockout reactions}
Knockout reactions of the form $A(p,pN)B$, in which an energetic proton collides with a target nucleus $A$, removing a nucleon $N$ and leaving a residual nucleus $B$, have been used as spectroscopic tools to study nucleon-hole states. The technique has recently experienced a revival thanks to the possibility of extending these studies to exotic nuclei, using reactions in inverse kinematics with a hydrogen target.  

\begin{figure}[htbp]
\begin{center}
\includegraphics[width=0.55\columnwidth]{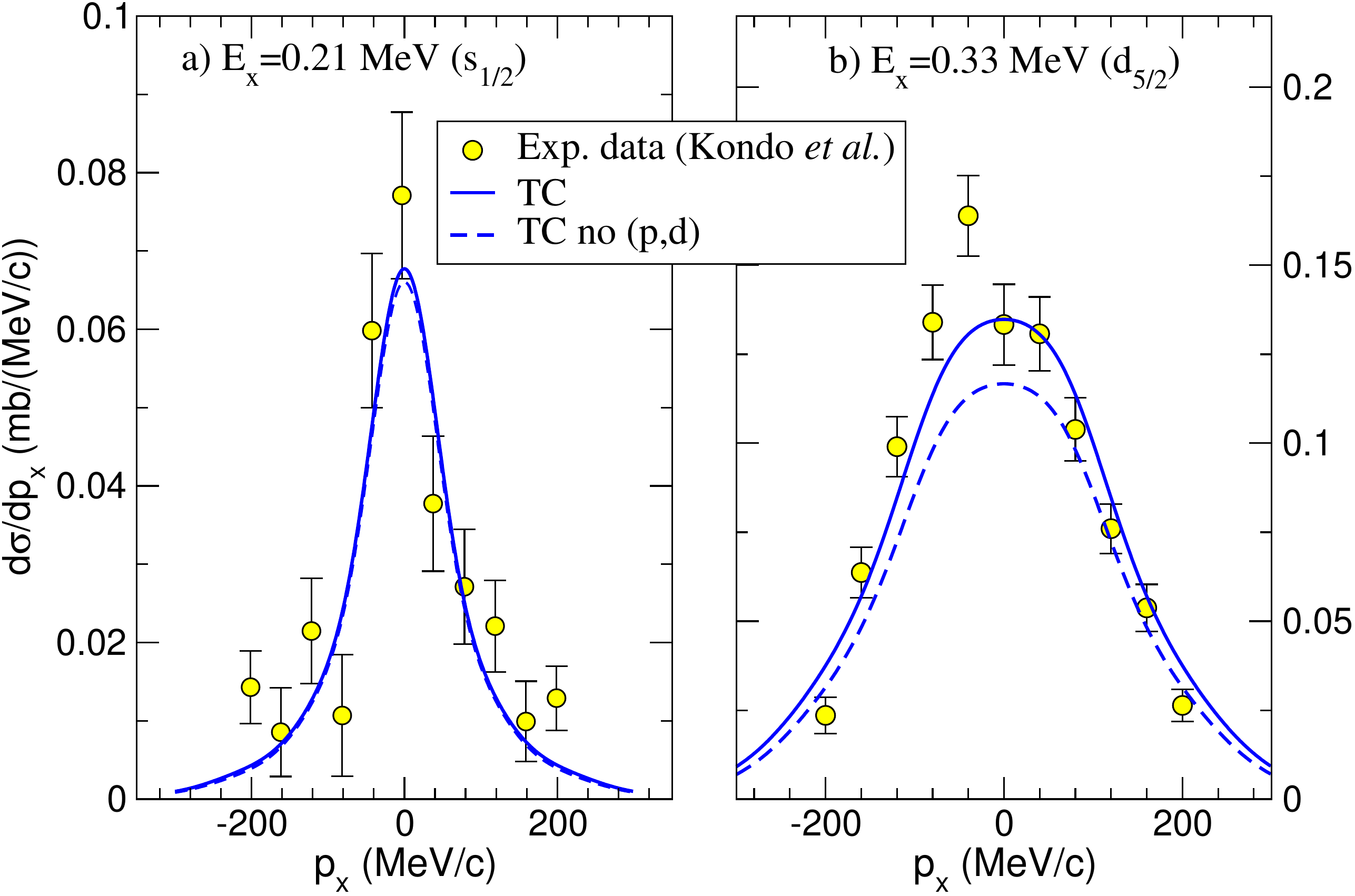}
\caption{\label{fig:c18ppn}Transverse $p_x$ momentum distributions for the $^{18}\text{C}(p, pn)^{17}\text{C}^*$
reaction. Experimental data are from \cite{Kon09}. The left and right panels correspond to the
population of the $^{17}$C states at $E_x = 0.21$~MeV and  $E_x = 0.33$~MeV. The blue solid line
corresponds to the full calculation rescaled to give the experimental total
cross section. The blue dashed line corresponds to the calculation removing
the contribution from the $(p, d)$ transfer reaction, rescaled by the same factor
as the full calculation. All theoretical calculations have been convoluted with
the experimental resolution.}
\end{center}
\end{figure}

Although the analysis of these reactions has been traditionally done with the DWIA method (see \cite{Wak17} for a recent review), some recent works have exploited alternative methods, such as the Faddeev/AGS equations \cite{Cre14,Cre16} or the so-called  {\it transfer-to-the-continuum} (TC) approach \cite{Mor15}. The latter is based on the prior-form of the transition amplitude for a process of the form $A(p,pN)B$, which can be written as 
\begin{align}
\mathcal{T}_{if}& =\left\langle \phi_{B}(\xi_B)  \Psi_{f}^{(-)}(\vec{r}_p,\vec{r}_N)
\Big| V_{pN}+U_{pB}-U_{pA} \Big| \Phi_A(\xi_A)  \chi_{pA}^{(+)}(\vec{R}) \right\rangle,
\end{align}
where $\Phi_A(\xi_A)$ is the ground-state wave function of the  projectile, $\phi_{B}(\xi_B)$ is the wavefunction of the residual nucleus $B$, $\Psi_{f}^{(-)}$ is the wavefunction describing the relative motion of the $p+N+B$ system and $\chi^{(+)}_{pA}$ is a distorted wave for $p+A$ scattering, obtained with some auxiliary potential $U_{pA}$. As in the case of $(d,p)$ reactions, the transition amplitude will be dominated by small $p$-$n$ separations, and hence the $\Psi_{f}^{(-)}$ can be approximated by a CDCC expansion in terms of $p$-$n$ states. Note that, in this approach, final-state interactions of the outgoing $p$-$n$ pair, including the deuteron bound state, are explicitly taken into account. Although the latter contribution is  expected to be small at sufficiently high energies, it can be important at lower energies (below 100~MeV/u), as demonstrated in Fig.~\ref{fig:c18ppn} for the $^{18}\text{C}(p, pn)^{17}\text{C}^∗$ reaction, where the left and right panels correspond to the population of two excited states of $^{17}$C. In particular, for the population of the excited state at $E_x=0.33$~MeV, the contribution of the $(p,d)$ is significant, and will therefore affect the extracted spectroscopic factor, since the latter is determined from the ratio of the experimental and theoretical crosss sections. 

\subsection{Transfer reactions populating unbound states}
\begin{figure}[htbp]
\begin{center}
\includegraphics[width=0.7\columnwidth]{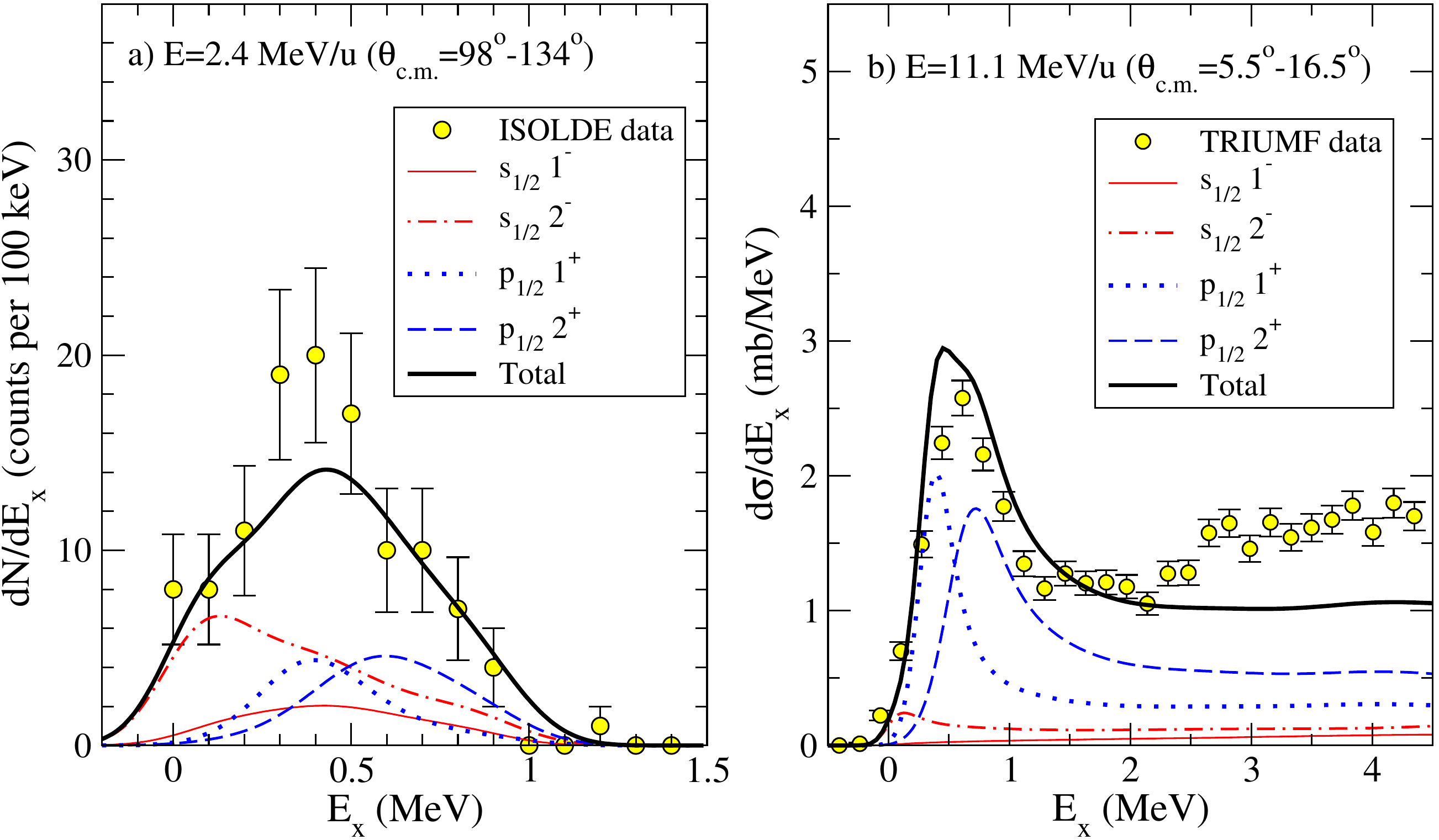}
\caption{\label{fig:li9dp} Differential cross section for $^{9}\text{Li}(d,p)^{10}\text{Li}$, as a function of the $^{10}$Li  excitation energy at an incident energy of 2.4~MeV/u (left) and 11.1~MeV/u (right). The data from Refs.~\cite{Jep06} and \cite{Cav17} are compared with TC calculations based on the same $^{10}$Li model. The separate contribution of $s$-waves ($1^-$, 2$^-$) and $p$-waves ($1^+$, $2^+$) are shown. Adapted from Ref.~\cite{Mor19}.}
\end{center}
\end{figure}

Transfer reactions studies with nuclei close to the driplines can be used to populate unbound systems and hence to investigate the evolution of nuclear structure properties, such as the shell ordering, beyond the limits of stability. An example is the $^{10}$Li system, which has been investigated with the  $^{9}$Li($d$,$p$)$^{10}$Li reaction at ISOLDE \cite{Jep06} and TRIUMF \cite{Cav17} facilities.  The transition amplitude for such process can be written, using the prior form representation, as
\begin{align}
\mathcal{T}_{if} &= \langle \Psi^{3b}_{\text{n-10Li}}|V_\text{n-9Li}+U_\text{p-9Li}-U_\text{d-9Li} | \phi_d(\vec{r}) \chi_{d-9\text{Li}}^{(+)}(\vec{R}) \rangle,
\label{eq:li9dp}
\end{align}
where the exact three-body wavefunction $\Psi^{3b}_{\text{n-10Li}}$ can be approximated by a CDCC expansion in $n$+$^{9}$Li states. Note that in this case no bound states are present in the expansion since the $^{10}$Li is unbound with respect to particle emission. In Fig.~\ref{fig:li9dp} we compare  the results \cite{Mor19} of the application of Eq.~(\ref{eq:li9dp}) to the $^{9}$Li($d$,$p$)$^{10}$Li reaction at 2.4~MeV/u and 11.1~MeV/u with the aforementioned data from ISOLDE and TRIUMF. Note that these differential cross sections are integrated over the angular range in which the outgoing particles were detected. The same structure model (i.e.\  $n$+$^{9}$Li interaction)  was employed at the two incident energies \cite{Cas17}. In this model, the low-lying $^{10}$Li continuum is characterized by a $s$-wave doublet ($1^-$, $2^-$), with a large effective scattering length,  and a  $p$-wave resonant doublet ($1^+$, $2^+$). Interestingly, these waves manifest themselves very differently  in the two reactions. In particular, for the higher incident energy, the contribution of  the $s$-wave virtual state very small and, in fact, it is not apparent in the data. As discussed in \cite{Mor19}, this seemingly different behaviour is mostly due to the different angular ranges considered in the two experiments.

\section{Application to non-elastic breakup and incomplete fusion}
The CDCC wavefunction is a projected few-body wavefunction defined in a modelspace in which the internal degrees of freedom of the considered bodies are not taken into account. Because of that, it describes explicitly only the so-called {\it elastic breakup} (EBU), that is, the projectile dissociation leaving the target in its ground state. The aforementioned extended versions of CDCC are able to include some {\it inelastic breakup} components, in which the projectile dissociation is accompanied by the excitation of the target or of the projectile fragments. In an inclusive experiment of the form $A(a,b)X$, in which only one of the outgoing fragments ($b$) is detected, there are however many other {\it non-elastic breakup} (NEB) components in which the unobserved particle ($x$) may interact in any possible way with the target nucleus. Although the CDCC wavefunction does not account explicitly for these NEB channels, their effect is actually embedded in the absorptive part of the fragment-target potentials. In the 1980s, Ichimura, Austern and Vincent (IAV) derived a closed-form formula for the evaluation of these NEB cross sections \cite{Ich85}. Originally, the model relied on the post-form DWBA formalism but, later on, Austern {\it et al.} \cite{Aus87} proposed a three-body version in which the $a+A$ distorted wave was replaced by a CDCC wavefunction. The IAV expression for the NEB differential cross section, as a function of the angle and energy of the detected fragment $b$ is given by:   
%
\begin{equation}
\label{eq:neb_iav}
\frac{d\sigma^\mathrm{NEB}}{d \Omega_b d E_b} = 
 - \frac{2}{\hbar v_a} \rho_b(E_b) \langle \varphi^{(\veck_{b})}_x | W_{xA} | \varphi^{(\veck_{b})}_x \rangle
\end{equation}
where $v_a$ is the projectile velocity, $\rho_b(E_b)$ the density of states of particle $b$,  $W_{xA}$the imaginary part of the $x-A$ optical potential, $\varphi^{(\veck_{b})}(\vecr_x)$ describes $x-A$ relative motion when $b$ scatters with momentum $\veck_{b}$. This function is calculated by solving the inhomogeneous equation
\begin{equation}
[K_x + U_{xA} -E_x] \varphi^{(\veck_{b})}_{x}(\vecr_x)  = -\langle \vecr_x \chi_b^{(-)}(\veck_{b}) | V_{bx} | \Psi^{(+)}_{3b} \rangle ,
\end{equation}
which contains the, in principle exact, three-body scattering wave function $\Psi^{(+)}_{3b}$. Austern {\it et al.}~\cite{Aus87} suggested approximating this function by the CDCC one (referred hereafter as IAV-CDCC formalism). 

\begin{figure}[htbp]
\begin{center}
\includegraphics[width=0.5\columnwidth]{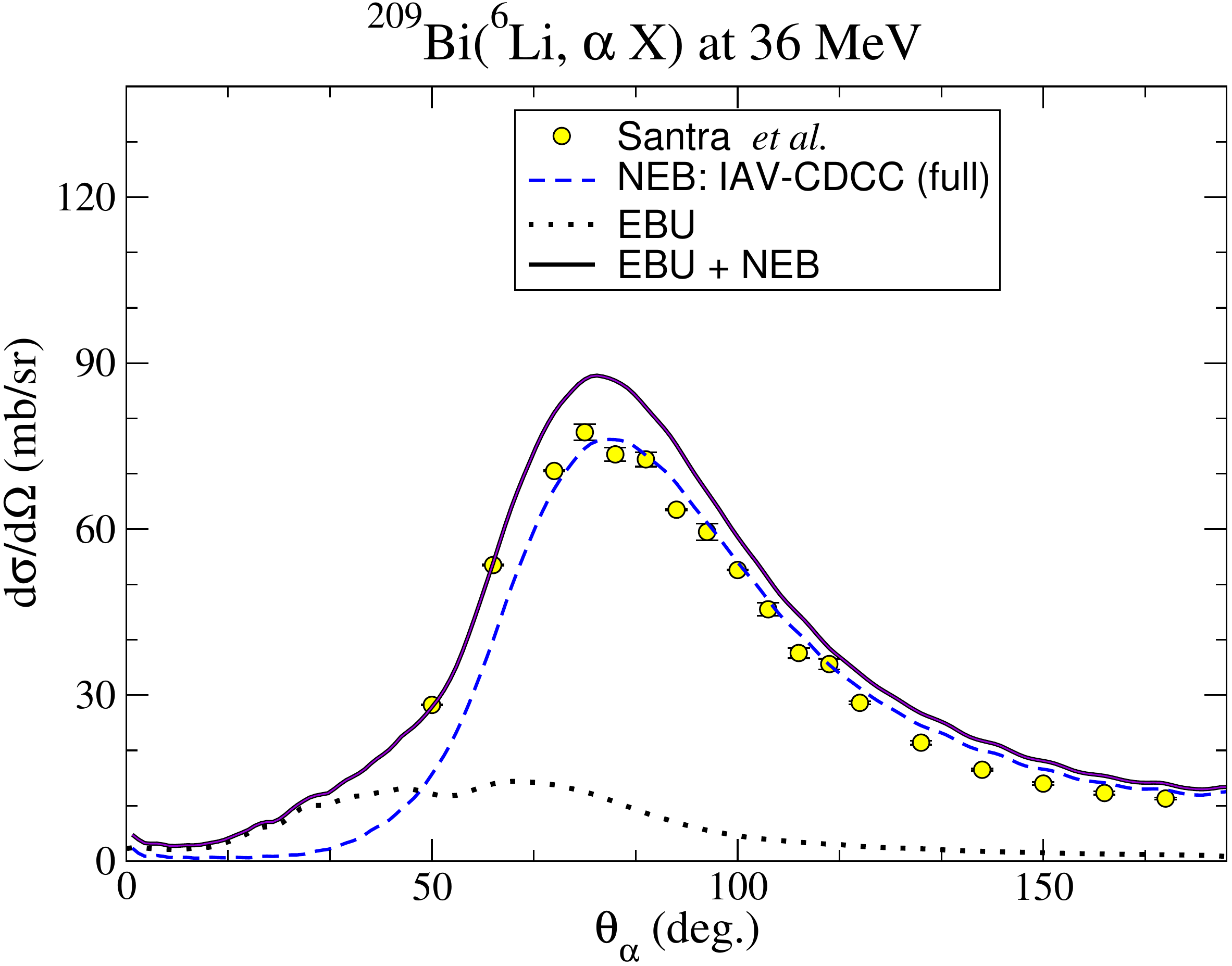}
\caption{\label{fig:li6bi_iav} Experimental and calculated $\alpha$ differential cross section resulting from the reaction  $^{6}$Li+$^{209}$Bi at 36 MeV. The data from Ref.~\cite{Santra12} are compared with calculations for the elastic breakup (EBU) and non-elastic breakup (NEB) components. See text for details.}. 
\end{center}
\end{figure}

The DWBA version of the IAV model has received renewed attention in recent years \cite{Jin15,Pot15,Car16}. Application of the IAV-CDCC model has been hampered due to its numerical complexity but, very recently, the first implementation of this model has been achieved and applied to some reactions \cite{Jin19}. An example is shown in Fig.~\ref{fig:li6bi_iav}, which corresponds to the reaction $^{209}$Bi($^{6}$Li,$\alpha$)X at an incident energy of 36~MeV. The inclusive breakup data are from Ref.~\cite{Santra12}. Elastic and non-elastic breakup contributions computed, respectively, with CDCC and IAV-CDCC, are shown as well as their sum. An interesting finding is that most of the measured inclusive cross section is due to NEB processes, whereas EBU represents a relatively small fraction and is only dominant at small angles. The full, EBU+NEB calculation,  reproduces rather well the shape and magnitude of the data, although some overestimation of the cross section at the maximum is visible.

\ack
This work has been partially supported by the National Science Foundation
under Contract No.\ NSF-PHY-1520972 with Ohio University,
by the Spanish Ministerio de Ciencia, Innovaci\'on y Universidades and FEDER funds under project FIS2017-88410-P  and by the European Union's Horizon 2020 research and innovation program under Grant Agreement No.\ 654002.

\section*{References}
\bibliographystyle{iopart-num}
\bibliography{refer}

\providecommand{\newblock}{}
\begin{thebibliography}{10}
\expandafter\ifx\csname url\endcsname\relax
  \def\url#1{{\tt #1}}\fi
\expandafter\ifx\csname urlprefix\endcsname\relax\def\urlprefix{URL }\fi
\providecommand{\eprint}[2][]{\url{#2}}

\bibitem{Raw74}
Rawitscher G~H 1974 {\em Phys. Rev. C\/} {\bf 9} 2210

\bibitem{Yah86}
Yahiro M, Iseri Y, Kameyama H, Kamimura M and Kawai M 1986 {\em Prog. Theor.
  Phys. Suppl.\/} {\bf 89} 32

\bibitem{Aus87}
Austern N, Iseri Y, Kamimura M, Kawai M, Rawitscher G and Yahiro M 1987 {\em
  Phys. Rep.\/} {\bf 154} 125

\bibitem{Kaw86a}
Kawai M 1986 {\em Prog. Theor. Phys. Supp.\/} {\bf 89} 11

\bibitem{Mor09b}
Moro A, Arias J, G{\'o}mez-Camacho J and P{\'e}rez-Bernal F 2009 {\em Phys.
  Rev. C\/} {\bf 80} 054605

\bibitem{Gom17a}
G\'omez-Ramos M and Moro A~M 2017 {\em Phys. Rev. C\/} {\bf 95} 034609

\bibitem{Sum06}
{Summers} N~C {\em et~al.\/} 2006 {\em Phys. Rev. C\/} {\bf 74} 014606

\bibitem{Die14}
{de Diego} R, {Arias} J~M, {Lay} J~A and {Moro} A~M 2014 {\em Phys. Rev. C\/}
  {\bf 89} 064609

\bibitem{Sum07}
{Summers} N~C and {Nunes} F~M 2007 {\em Phys. Rev. C\/} {\bf 76} 014611

\bibitem{Mor12b}
Moro A~M and Lay J~A 2012 {\em Phys. Rev. Lett.\/} {\bf 109} 232502

\bibitem{Pes17}
Pesudo V {\em et~al.\/} 2017 {\em Phys. Rev. Lett.\/} {\bf 118} 152502

\bibitem{Chen16a}
Chen J, Lou J, Ye Y, Li Z, Ge Y, Li Q, Li J, Jiang W, Sun Y, Zang H {\em
  et~al.\/} 2016 {\em Phys. Rev. C\/} {\bf 93} 034623

\bibitem{Chen16b}
Chen J, Lou J, Ye Y, Rangel J, Moro A, Pang D, Li Z, Ge Y, Li Q, Li J {\em
  et~al.\/} 2016 {\em Phys. Rev. C\/} {\bf 94} 064620

\bibitem{Dip19}
Pietro A~D, Moro A~M, Lei J and de~Diego R 2019 {\em Phys. Lett. B\/} {\bf 798}
  134954

\bibitem{Lay16}
Lay J, de~Diego R, Crespo R, Moro A, Arias J and Johnson R~C 2016 {\em Phys.
  Rev. C\/} {\bf 94} 021602

\bibitem{Pan13}
Pang D~Y, Timofeyuk N~K, Johnson R~C and Tostevin J~A 2013 {\em Phys. Rev. C\/}
  {\bf 87} 064613

\bibitem{Cha17}
Chazono Y, Yoshida K and Ogata K 2017 {\em Phys. Rev. C\/} {\bf 95} 064608

\bibitem{Ron70}
Johnson R~C and Soper P~J~R 1970 {\em Phys. Rev. C\/} {\bf 1} 976

\bibitem{JT74}
Johnson R~C and Tandy P~C 1974 {\em Nucl. Phys. A\/} {\bf 235} 56 ISSN
  0375-9474

\bibitem{Gom17b}
Gomez-Ramos M and Moro A~M 2017 {\em Phys. Rev. C\/} {\bf 95}(4) 044612

\bibitem{Wak17}
Wakasa T, Ogata K and Noro T 2017 {\em Progress in Particle and Nuclear
  Physics\/} {\bf 96} 32

\bibitem{Cre14}
Crespo R, Deltuva A and Cravo E 2014 {\em Phys. Rev. C\/} {\bf 90} 044606

\bibitem{Cre16}
Cravo E, Crespo R and Deltuva A 2016 {\em Phys. Rev. C\/} {\bf 93} 054612

\bibitem{Mor15}
Moro A~M 2015 {\em Phys. Rev. C\/} {\bf 92}(4) 044605

\bibitem{Kon09}
Kondo Y {\em et~al.\/} 2009 {\em Phys. Rev. C\/} {\bf 79} 014602

\bibitem{Jep06}
Jeppesen H~B {\em et~al.\/} 2006 {\em Phys. Lett. B\/} {\bf 642} 449

\bibitem{Cav17}
Cavallaro M {\em et~al.\/} 2017 {\em Phys. Rev. Lett.\/} {\bf 118} 012701

\bibitem{Mor19}
Moro A, Casal J and G\'omez-Ramos M 2019 {\em Physics Letters B\/} {\bf 793} 13

\bibitem{Cas17}
Casal J, G\'omez-Ramos M and Moro A~M 2017 {\em Phys. Lett. B\/} {\bf 767} 307

\bibitem{Ich85}
Ichimura M, Austern N and Vincent C~M 1985 {\em Phys. Rev. C\/} {\bf 32} 431

\bibitem{Jin15}
Lei J and {Moro} A~M 2015 {\em Phys. Rev. C\/} {\bf 92} 044616

\bibitem{Pot15}
Potel G, Nunes F~M and Thompson I~J 2015 {\em Phys. Rev. C\/} {\bf 92} 034611

\bibitem{Car16}
Carlson B~V, Capote R and Sin M 2016 {\em Few-Body Systems\/} {\bf 57} 307

\bibitem{Jin19}
Lei J and {Moro} A~M 2019 {\em Submitted\/}

\bibitem{Santra12}
Santra S, Kailas S, Parkar V~V, Ramachandran K, Jha V, Chatterjee A, Rath P~K
  and Parihari A 2012 {\em Phys. Rev. C\/} {\bf 85} 014612

\end{thebibliography}

\end{document}